\documentclass[aps,preprint,showpacs]{revtex4}
\begin{document}
\title{The fundamental spherically symmetric fluid model}
\author{B.V.Ivanov}
\email{boyko@inrne.bas.bg}
\affiliation{Institute for Nuclear Research and Nuclear Energy, Bulgarian Academy of\\
Sciences, Tzarigradsko Shausse 72, Sofia 1784, Bulgaria}

\begin{abstract}
It is shown that an effective anisotropic spherically symmetric fluid model
with heat flow can absorb the addition to a perfect fluid of pressure
anisotropy, heat flow, bulk and shear viscosity, electric field and null
fluid. In most cases the induction of effective heat flow can be avoided.
There is a relationship between anisotropic and charged perfect fluids.
\end{abstract}
\pacs{04.20 Cv, 04.20 Dg, 04.40 Nr}
\maketitle
\section{Introduction}

Spherically symmetric perfect fluid solutions in general relativity have
been studied from its very beginning, starting with the interior
Schwarzschild solution. Gradually different mechanisms in stellar models
have been identified that create pressure anisotropy \cite{one}, \cite{two}.
Radiating spherical collapse demands the introduction of heat flow \cite
{three} or null fluid which describe energy dissipation in different
approximations \cite{four}. More realistic fluids possess also bulk and
shear viscosity \cite{five}, \cite{six}, \cite{seven}. Charged perfect
fluids or dust have been discussed by many authors \cite{eight}. It has been
shown too that the sum of two perfect fluids, two null fluids or a perfect
and a null fluid can be represented by effective anisotropic fluid models 
\cite{nine}. In view of the many existing relations among the fluid models
mentioned above it is hard to point out some center.

In this paper it is shown that the anisotropic fluid with heat flow is in
some sense the most fundamental model and can absorb the addition of
viscosity, charge and null fluids. Something more, heat flow is not
generated in most cases.

In Sec.II the basic anisotropic fluid model is defined. In Sec.III the fluid
is supplied with bulk and shear viscosity which leads to a new effective
anisotropic model. In Sec.IV the same is done for the addition of charge and
in Sec.V null fluids are accommodated into the anisotropic model. Sec.VI
summarizes all additions and the effective characteristics of the general
anisotropic model are given. Several conclusions are drawn.

\section{Anisotropic fluid model with heat flow}

Einstein's field equations are given by 
\begin{equation}
8\pi T_{\alpha \beta }=G_{\alpha \beta }  \label{one}
\end{equation}
where $G_{\alpha \beta }$ is the Einstein tensor, $T_{\alpha \beta }$ is the
energy-momentum tensor (EMT) and units are used so that $c=G=1$. The general
spherically symmetric metric is written as 
\begin{equation}
ds^2=-A^2dt^2+B^2dr^2+R^2\left( d\theta ^2+\sin ^2\theta \ d\varphi ^2\right)
\label{two}
\end{equation}
where $A,B,R$ are positive functions of $t$ and $r$ only. The spherical
coordinates are numbered as $x^0=t,x^1=r,x^2=\theta ,x^3=\varphi $. The
Einstein tensor involves the Ricci tensor and scalar which are given by the
metric and its first and second derivatives \cite{four}. Its non-trivial
components are $G_{00},G_{01},G_{11},G_{22}=\sin ^{-2}\theta \ G_{33}.$

We are interested in the structure of EMT. It reads for anisotropic fluids
with heat flow 
\begin{equation}
T_{\alpha \beta }=\left( \mu +p_t\right) u_\alpha u_\beta +p_tg_{\alpha
\beta }+\left( p_r-p_t\right) \chi _\alpha \chi _\beta +q_\alpha u_\beta
+u_\alpha q_\beta .  \label{three}
\end{equation}
Here $\mu $ is the energy density, $p_r$ is the radial pressure, $p_t$ is
the tangential pressure, $u^\alpha $ is the four velocity of the fluid (a
timelike vector), $\chi ^\alpha $ is a unit spacelike vector along the
radial direction and $q^\alpha $ is the heat flux (in the radial direction
too). We have 
\begin{equation}
u^\alpha u_\alpha =-1,\quad \chi ^\alpha \chi _\alpha =1,\quad u^\alpha \chi
_\alpha =0,\quad u^\alpha q_\alpha =0.  \label{four}
\end{equation}
It is assumed that the coordinates are comoving, hence, the fluid is
motionless in them 
\begin{equation}
u^\alpha =A^{-1}\delta _0^\alpha ,\quad \chi ^\alpha =B^{-1}\delta _1^\alpha
,\quad q^\alpha =qB^{-1}\delta _1^\alpha  \label{five}
\end{equation}
where $q=q\left( r,t\right) $. This gives 
\begin{equation}
T_{00}=\mu A^2,\quad T_{01}=-qAB,\quad T_{11}=p_rB^2,\quad T_{22}=p_tR^2,
\label{six}
\end{equation}
which should be plugged in the Einstein equations. The anisotropic fluid
does not radiate when $q=0$ and becomes perfect when $p_r=p_t$. Thus it
accommodates anisotropy of pressure and heat flow when one starts with a
perfect fluid model.

Let us see now what happens when other EMTs are added to the basic
anisotropic one.

\section{Bulk and shear viscosity}

Bulk viscosity \cite{five}, \cite{six}, \cite{seven} adds to the basic EMT
the following piece 
\begin{equation}
T_{\alpha \beta }^B=-\zeta \Theta h_{\alpha \beta }  \label{seven}
\end{equation}
where $\zeta $ is a coefficient, $\Theta $ is the expansion of the fluid and 
$h_{\alpha \beta }$ is the projector on the hyperplane orthogonal to $%
u^\alpha $%
\begin{equation}
\Theta =u_{\;;\alpha }^\alpha ,\quad h_{\alpha \beta }=g_{\alpha \beta
}+u_\alpha u_\beta .  \label{eight}
\end{equation}
Obviously, this means the appearance of effective pressures 
\begin{equation}
p_r^B=p_t^B=-\zeta \Theta ,  \label{nine}
\end{equation}
which should be added to $p_r,p_t$. They do not change the degree of
anisotropy $\Delta p=p_r-p_t$. Thus even perfect fluid can absorb the bulk
viscosity. The quantities $\mu ,q$ remain the same. No heat flow is
generated in particular.

Shear viscosity is responsible for the piece 
\begin{equation}
T_{\alpha \beta }^S=-2\eta h_{\alpha \gamma }h_{\beta \delta }\sigma
^{\gamma \delta }  \label{ten}
\end{equation}
where $\sigma _{\alpha \beta }$ is the shearing tensor 
\begin{equation}
\sigma _{\alpha \beta }=u_{\left( \alpha ;\beta \right) }+a_{(\alpha
}u_{\beta )}-\frac 13\Theta h_{\alpha \beta },  \label{eleven}
\end{equation}
$\eta $ is some coefficient and $a_\alpha $ is the acceleration 
\begin{equation}
a_\alpha =u_{\alpha ;\beta }u^\beta .  \label{twelve}
\end{equation}
The shearing tensor satisfies the conditions 
\begin{equation}
\sigma _{\alpha \beta }u^\beta =0,\quad \sigma _{\alpha \beta }g^{\alpha
\beta }=0,  \label{thirteen}
\end{equation}
hence, Eq (10) transforms into 
\begin{equation}
T_{\alpha \beta }^S=-2\eta \sigma _{\alpha \beta }.  \label{fourteen}
\end{equation}
Use of Eqs (5,11) gives the non-zero components of the shear \cite{four} 
\begin{equation}
\sigma _{11}=\frac 23B^2\sigma ,\quad \sigma _{22}=\frac{\sigma _{33}}{\sin
^2\theta }=-\frac 13R^2\sigma  \label{fifteen}
\end{equation}
where 
\begin{equation}
\frac 23\sigma ^2=\sigma ^{\alpha \beta }\sigma _{\alpha \beta }.
\label{sixteen}
\end{equation}
One can check that the same components follow when $\sigma _{\alpha \beta }$
is written as the tensor 
\begin{equation}
\sigma _{\alpha \beta }=-\frac 13\sigma h_{\alpha \beta }+\sigma \chi
_\alpha \chi _\beta .  \label{seventeen}
\end{equation}
Thus it coincides with the general shear tensor defined by Eq (11) in the
spherically symmetric case. It also satisfies relations (13) in any metric.

Plugging Eq (17) into Eq (14) and comparing it to Eq (3) we find the
effectively generated pressures 
\begin{equation}
p_r^S=-2p_t^S=-\frac 43\eta \sigma .  \label{eighteen}
\end{equation}
The degree of anisotropy is changed. There is no generation of energy
density or heat flow. The scalars $\Theta ,\sigma $ can be expressed through
the metric and its first derivatives.

\section{Electromagnetic fields}

The EMT of electromagnetic fields is given by 
\begin{equation}
T_{\alpha \beta }^{EM}=\frac 1{4\pi }\left( F_{\mu \alpha }F_{\;\beta }^\mu -%
\frac 14g_{\alpha \beta }F_{\mu \nu }F^{\mu \nu }\right)  \label{nineteen}
\end{equation}
where $F_{\mu \nu }$ is the Faraday tensor. One defines a unit timelike
vector field $n^\mu $. An observer moving on its direction will measure
electric and magnetic field respectively 
\begin{equation}
E_\alpha =F_{\alpha \mu }n^\mu ,\quad H_\alpha =\frac 12\varepsilon _{\alpha
\mu \nu }F^{\mu \nu }.  \label{twenty}
\end{equation}
These fields are spacelike, $E^\alpha n_\alpha =H^\alpha n_\alpha =0$. The
Faraday tensor decomposes like \cite{ten} 
\begin{equation}
F_{\alpha \beta }=\varepsilon _{\alpha \beta \mu }H^\mu -2E_{[\alpha
}n_{\beta ]}.  \label{twone}
\end{equation}
Plugging this expression into Eq (19) gives formula (7) from Ref \cite{ten},
which becomes after some rearrangements 
\begin{equation}
T_{\alpha \beta }^{EM}=\frac 1{4\pi }\left( H^2+E^2\right) \left( n_\alpha
n_\beta +\frac 12g_{\alpha \beta }\right) -\frac 1{4\pi }\left( E_\alpha
E_\beta +H_\alpha H_\beta \right) +2j_{(\alpha }n_{\beta )}  \label{twtwo}
\end{equation}
where $E^2=E_\mu E^\mu ,H^2=H_\mu H^\mu $ and $j_\alpha $ is the Poynting
vector that measures the energy flow in the spacetime 
\begin{equation}
j_\alpha =\frac 1{4\pi }\varepsilon _{\alpha \mu \nu }E^\mu H^\nu .
\label{twthree}
\end{equation}

In order to absorb this EMT by the EMT for anisotropic fluid we choose the
direction $n^\alpha =u^\alpha $ and $\chi ^\alpha =E^\alpha /E$. The latter
is possible because when spherical symmetry is imposed $H_\alpha =0$ and $%
E_\alpha $ has only a radial spatial component. The would be heat flow term
in Eq (22) disappears and we get 
\begin{equation}
T_{\alpha \beta }^E=2eu_\alpha u_\beta +eg_{\alpha \beta }-2e\chi _\alpha
\chi _\beta ,\quad e=\frac{E^2}{8\pi }.  \label{twfour}
\end{equation}
Thus the addition of electric field induces effective pressures and energy
density 
\begin{equation}
\mu ^E=p_t^E=-p_r^E=e  \label{twfive}
\end{equation}
related by simple linear equations of state. Hence, a charged perfect fluid
or a charged anisotropic fluid may be represented effectively by some
neutral anisotropic fluid. There is no heat flow induction in this case.

\section{Null fluid}

Null fluid describes dissipation in the free streaming approximation \cite
{four} and adds to the basic EMT the piece 
\begin{equation}
T_{\alpha \beta }^N=\varepsilon l_\alpha l_\beta  \label{twsix}
\end{equation}
where $l^\alpha $ is the null vector 
\begin{equation}
l^\alpha =A^{-1}\delta _0^\alpha +B^{-1}\delta _1^\alpha =u^\alpha +\chi
^\alpha ,  \label{twseven}
\end{equation}
satisfying the relations 
\begin{equation}
l^\mu l_\mu =0,\quad l^\mu u_\mu =-1.  \label{tweight}
\end{equation}
Substituting Eq (27) into Eq (26) one finds 
\begin{equation}
T_{\alpha \beta }^N=\varepsilon u_\alpha u_\beta +\varepsilon \chi _\alpha
\chi _\beta +\varepsilon \left( u_\alpha \chi _\beta +u_\beta \chi _\alpha
\right) .  \label{twnine}
\end{equation}
A comparison of this expression with Eq (3), taking into account that $%
q^\alpha =q\chi ^\alpha $, shows that the addition of null fluid generates
effective energy density, radial pressure and heat flow, all of them equal 
\begin{equation}
\mu ^N=p_r^N=q^N=\varepsilon .  \label{thirty}
\end{equation}
No tangential pressure is generated. This is the only case where an
effective heat flow is induced.

On the other hand, it is known \cite{nine}, \cite{eleven} that a perfect
fluid plus a null fluid are equivalent to an anisotropic fluid without a
heat flow when a rotation of $u^\alpha ,l^\alpha $ is done. This is true in
any metric. Can we do the same for the sum of anisotropic and null fluid?
The Letelier method is based on the condition $\tilde u^\mu \tilde l_\mu =0$%
, imposed on the rotated vectors. Then one of them is necessarily timelike,
the other spacelike and the EMT of the sum may be written as in Eq (3) with $%
q=0$. However, when the basic model is anisotropic, there is already a
spacelike vector, namely $\chi ^\alpha $. Therefore we shall rotate $%
l^\alpha $ into it. Following Ref. \cite{nine} we change the signature of
the metric and extract from $T_{\alpha \beta }+T_{\alpha \beta }^N$ the part 
\begin{equation}
\left( \mu +p_t\right) u_\alpha u_\beta +\varepsilon l_\alpha l_\beta
\label{thone}
\end{equation}
where now $u_\mu u^\mu =1,\chi ^\mu \chi _\mu =-1.$ The above expression
remains invariant under a rotation with angle $\alpha $%
\begin{equation}
\tilde u^\alpha =\cos \alpha \ u^\alpha +C\sin \alpha \ l^\alpha ,\quad 
\tilde l^\alpha =-C^{-1}\sin \alpha \ u^\alpha +\cos \alpha \ l^\alpha ,
\label{thtwo}
\end{equation}
\begin{equation}
C=\left( \frac \varepsilon {\mu +p_t}\right) ^{1/2}.  \label{ththree}
\end{equation}
Let us demand that $\tilde l^\alpha $ is proportional to $\chi ^\alpha $ and
thus is spacelike 
\begin{equation}
\tilde l^\alpha =a\chi ^\alpha ,\quad \tilde l^\alpha \tilde l_\alpha =-a^2.
\label{thfour}
\end{equation}
The relation $\chi ^\mu u_\mu =0$ yields 
\begin{equation}
\tan \alpha =Al^\mu u_\mu ,\quad a^2=\frac{\sin ^2\alpha }{A^2}.
\label{thfive}
\end{equation}
Next we find 
\begin{equation}
\tilde u^\mu \tilde u_\mu =1+\sin ^2\alpha ,\quad \tilde u^\mu \tilde l_\mu
=-\frac{\sin ^2\alpha }Au^\mu l_\mu .  \label{thsix}
\end{equation}
Thus $\tilde u^\alpha $ is timelike but is not orthogonal to the spacelike $%
\tilde l^\alpha $. The EMT of the two fluids sum reads 
\begin{equation}
\left( \mu +p_t\right) \tilde u_\alpha \tilde u_\beta -p_tg_{\alpha \beta
}+\left( \frac{p_r-p_t}{a^2}+\varepsilon \right) \tilde l_\alpha \tilde l%
_\beta .  \label{thseven}
\end{equation}
Now we normalize $\tilde u^\alpha ,\tilde l^\alpha $ and then perform
another Letelier rotation to make them orthogonal, $\hat u^\mu \hat l_\mu =0$%
. This can be done when $p_r-p_t+\varepsilon a^2>0$. This holds e.g. when $%
p_r>p_t$. As a result $\hat u^\alpha $ stays timelike, while $\hat l^\alpha $
stays spacelike and the EMT becomes that of anisotropic fluid without a heat
flow. The expressions for the effective energy density and pressures are
more complicated than the lengthy formulas in \cite{nine} because two
rotations have been performed, so we omit them here.

\section{Summary and conclusions}

The results in the previous sections show that viscosity, electric charge
and null fluids are equivalent to induced energy density and pressures,
related by simple linear equations of state $\mu =np_r$, $n=0,\pm 1$ and $%
p_t=kp_r$, $k=0,\pm 1,-1/2$. When all such additions are combined and
absorbed one obtains an effective anisotropic fluid model with 
\begin{equation}
\mu ^e=\mu +e+\varepsilon ,  \label{theight}
\end{equation}
\begin{equation}
p_r^e=p_r-\zeta \Theta -\frac 43\eta \sigma -e+\varepsilon ,  \label{thnine}
\end{equation}
\begin{equation}
p_t^e=p_t-\zeta \Theta +\frac 23\eta \sigma +e+\varepsilon ,  \label{forty}
\end{equation}
\begin{equation}
q^e=q+\varepsilon .  \label{foone}
\end{equation}
These should be plugged into Eq (6) and hence in the l.h.s. of the Einstein
equations (1). There are 4 equations for 8 functions ($\mu ,p_r,p_t,q,\zeta
,\eta ,e,\varepsilon $). The quantities $\sigma ,\Theta $ are expressed
through the metric. One has to impose 4 relations on these functions or set
some of them to zero in order to obtain a determined system of equations.
Several conclusions can be drawn.

Only the functions of the two modes of dissipation of energy ($q,\varepsilon 
$) have effect upon the heat flow.

Viscosity ($\zeta ,\eta $) does not induce effective energy density.

An important characteristic is the anisotropy factor 
\begin{equation}
\bigtriangleup p^e=p_r^e-p_t^e=\bigtriangleup p-2\eta \sigma -2e.
\label{fotwo}
\end{equation}
We see that shear viscosity and charge induce pressure anisotropy. Their
absorption by a perfect fluid ($\bigtriangleup p=0$) makes the latter
anisotropic and adds to more sources of anisotropy to the usual ones \cite
{one}.

Bulk viscosity and null fluid don't induce anisotropy and may be absorbed
into an effective perfect fluid model with heat flow.

Charged perfect fluids are related to anisotropic fluids by 
\begin{equation}
\bigtriangleup p=-2e.  \label{fothree}
\end{equation}
In addition to the Einstein equations charged fluids have a Maxwell equation
but it serves as a definition of the charge $Q$ in terms of $e$. Thus
results about anisotropic fluids may be carried over to charged perfect
fluids. For example, in Ref \cite{twelve} all static spherically symmetric
anisotropic fluid solutions are generated by two arbitrary functions, $Z$
and $\bigtriangleup p$. When $\bigtriangleup p=0$ their formula reduces to
the result for perfect fluids \cite{thirteen}. When Eq (43) holds instead,
we obtain a description of charged perfect fluids in terms of $Z$ and $e$.

\end{document}